# Many-body and spin-orbit effects on direct-indirect band gap transition of strained monolayer MoS$_2$ and WS$_2$

Luqing Wang,[1] Alex Kutana,[1] and Boris I. Yakobson[1,2,3,*]

[1]*Department of Materials Science and Nanoengineering,* [2]*Department of Chemistry, and* [3]*Smalley Institute for Nanoscale Science and Technology, Rice University, Houston, Texas 77005, USA*

*e-mail: biy@rice.edu phone: (713)348-3572 fax: (713)348-5423

**Abstract**

Monolayer transition metal dichalcogenides are promising materials for photoelectronic devices. Among them, molybdenum disulphide (MoS$_2$) and tungsten disulphide (WS$_2$) are some of the best candidates due to their favorable band gap values and band edge alignments. Here we consider various perturbative corrections to the DFT electronic structure, e.g. GW, spin-orbit coupling, as well as many-body excitonic and trionic effects, and calculate accurate band gaps as a function of homogeneous biaxial strain in these materials. We show that all of these corrections are of comparable magnitudes and need to be included in order to obtain an accurate electronic structure. We calculate the strain at which the direct-to-indirect gap transition occurs. After considering all contributions, the direct to indirect gap transition strain is predicted to be at 2.7% in MoS$_2$ and 3.9% in WS$_2$. These values are generally higher than the previously reported theoretical values.

Strain tuning of electronic and optical properties of materials could provide a new approach to enhancing their characteristics, e.g. increasing carrier mobilities or adjusting electronic band gaps. 2D materials represent an important group where strain tuning can be easily achieved. [1] In particular, in monolayer MoS$_2$ and WS$_2$ strain tuning may be used to vary the band gaps and induce a direct to indirect band gap transition. The band gap values and band edge positions in these two-dimensional transition metal dichalcogenides (TMDs) are well suited for solar energy harvesting and photoelectric applications. Unlike their bulk counterparts, unstrained monolayer MoS$_2$ and WS$_2$ have direct gaps, and undergo a direct-indirect gap transition under applied strain. As the indirect band gap may limits their optical uses, the critical strain at which the gap transition occurs in these materials must be known accurately; however, it is not yet firmly established either experimentally [2-4], or theoretically. [5,6,7]

Conley *et al*. [2] reported the direct-to-indirect transition in the optical band gap of monolayer MoS$_2$ at the applied strain of 1%, based on the decrease in the photoluminescence intensity. However, peaks are assigned by deconvolution, which is not always reliable, and the decrease is also observed at lower strains, e.g. 0.5% or 0.6%, so that in fact there is no sharp change indicating the transition. No indirect transition peak is observed at 1%, which would be indicative of the creation of the indirect gap. He *et al*. [3] and Gomez *et al*. [4] observed the strain-induced reduction of both the direct and indirect band gaps by photoluminescence, although the exact transition point to the indirect gap was not reported. Previously, theoretical studies of the effect of strain on electronic structure in MoS$_2$ and WS$_2$ TMDs have been performed, often yielding differing results.

A number of theoretical studies [5] have concluded that the gap transition in MoS$_2$ occurs at 1% strain based on the PBE calculations. These calculations considered neither the spin-orbit coupling nor many-body effects. Zhang *et al*. [7] have found that the monolayer MoS$_2$ remains a direct gap semiconductor up to 4% uniaxial strain with inclusion of spin-orbit coupling, but without the consideration of excitonic effects. Theoretical results of Conley *et al*. [2] show the gap transition at 5% strain based on the GW approximation, and at 0.1% strain when including excitonic effect, while GW calculations by Shi *et al*. [6] give the strain of 1% without adding the spin-orbit coupling. These discrepancies in calculated transition strains indicate that many significant contributions (spin-orbit coupling, excitonic, trionic, and other types of electron correlations, electron-phonon interactions) affect the value of the band gap and its dependence on strain. In this work, we evaluate contributions from the several (leading) of these effects and show that they indeed have comparable magnitudes. We next include these contributions to obtain the most accurate theoretical band gaps and transition strains in these materials. We also explain the differences that exist in the reported values of the band gaps that have been calculated earlier.

We use the density functional theory (DFT) with the PBE functional as an unperturbed solution, and apply various corrections, such as spin-orbit coupling (SOC) and GW formalism to obtain self-energies and modified Kohn-Sham energy eigenvalues, and also consider many-body excitonic and trionic effects. In line with the general treatment of the first-order perturbation theory, each of the effects is considered independently, and the corrections are then added up to yield the final values of energies and band gaps.

DFT calculations were performed within both the local density approximation (LDA) and generalized gradient approximation (GGA, PBE functional) to the exchange-correlation potential, using the projector augmented wave (PAW) method, as implemented in the Vienna ab initio simulation package (VASP). [8] Twelve valence electrons are included for Mo and W pseudo-atoms, and six electrons for the S pseudo-atom. The electronic wave function was expanded in a plane wave basis set with the kinetic energy cutoff of 400 eV. A vacuum slab of more than 20 Å (periodical length in $z$ direction is 24 Å) is added in the direction normal to the nano-sheet plane. For the Brillouin zone integration, a 9 × 9 × 1 Monkhorst-Pack $k$-point mesh is used. GW calculations were based on the PBE solutions. We employed 84 empty bands to calculate the full frequency-dependent dielectric function and obtain the G$_0$W$_0$ quasiparticle energies. The spin-orbit coupling was included into our calculations using a PAW implementation in VASP.

Exciton and trion binding energies were obtained within the effective mass model, using the potential proposed by Keldysh for thin films. [9] The effective in-plane 2D interaction for the charges separated by the distance $\rho = (x^2 + y^2)^{1/2}$ has the form: [9]

$$V_{2D}(\rho) = \frac{\pi e^2}{(\varepsilon_1+\varepsilon_2)\rho_0}\left[H_0\left(\frac{\rho}{\rho_0}\right) - Y_0\left(\frac{\rho}{\rho_0}\right)\right] \qquad (1)$$

where $H_0$ and $Y_0$ are the Struve function and the Bessel function of the second kind, and $\varepsilon_1$ and $\varepsilon_2$ are the permittivities of the surrounding media. In the strictly 2D limit of a polarizable plane in vacuum ($\varepsilon_{1,2} = 1$), the screening length is given by $\rho_0 = 2\pi\chi_{2D}$, where $\chi_{2D}$ is the 2D polarizability of the planar material. [10] To extract the 2D polarizability and thus the screening length $\rho_0$, we employ the relation: [11]

$$\varepsilon^{xy}(L_c) = 1 + \frac{4\pi\chi_{2D}}{L_c} + O\left(\frac{1}{L_c^2}\right) \qquad (2)$$

where $L_c$ is the interlayer separation for a supercell containing two AB-aligned monolayers of MoS$_2$ or WS$_2$ separated by vacuum. The in-plane dielectric constant $\varepsilon^{xy}$ is the $(q_x, q_y) \to 0$ limit (head) of the full dielectric function. Expressing the reduced mass $\mu$ as $1/\mu = 1/m_e + 1/m_h$, where $m_e$ and $m_h$ are the electron and hole effective masses, respectively, the trion Hamiltonian can be written as: [11]

$$H_{tri} = -\frac{1}{2\mu}\left(\nabla^2_{\rho_1} + \nabla^2_{\rho_2}\right) - \frac{1}{2m_h}\nabla_{\rho_1}\cdot\nabla_{\rho_2} - V_{2D}(\rho_1) - V_{2D}(\rho_2) + V_{2D}(|\rho_1 - \rho_2|) \qquad (3)$$

For the trion envelope wave function, the following variational form is considered: [12]

$$\psi_{tri}(\rho_1,\rho_2;a,b) = 2^{-1/2}[\psi_{ex}(\rho_1;a)\psi_{ex}(\rho_2;b) + \psi_{ex}(\rho_1;b)\psi_{ex}(\rho_2;a)] \qquad (4)$$

Here, $\psi_{ex}(\rho;a) = a(2/\pi)^{1/2}\exp(-\rho a)$. Strain-less monolayer MoS$_2$ and WS$_2$ are a K to K direct band gap semiconductors with respective experimental optical band gap of ~1.85 eV and ~1.99 eV. [13] A transition from a K to K direct band gap to a $\Gamma$ to K indirect band gap can be induced by strain in these materials. As the strain increases, the highest filled band at the $\Gamma$ point goes up and overtakes the maximum at the K point as the top of the valence band, resulting in the transition. The calculated electronic band structures of monolayer WS$_2$ at various biaxial strains are shown in Fig. 1 in three approximations: PBE without and with spin-orbit coupling (PBE+SOC), and G$_0$W$_0$ approximation based on the PBE solution without SOC (PBE+GW). Band structures are shown for 0%, 1%, and 5% strain. One notices appreciable differences among these solutions, and that corrections given to DFT by including SOC and performing GW are about same order of magnitude.

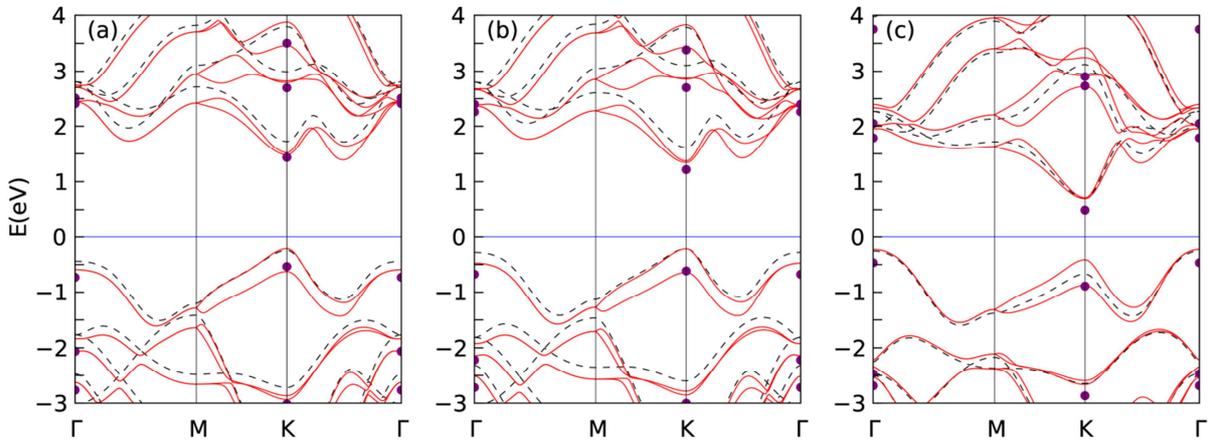

*Figure 1. PBE, PBE+SOC, and GW band structures of monolayer WS$_2$ under biaxial strain of (a) 0%, (b) 1% and (c) 5%, respectively. PBE bands are shown by dashed black lines, PBE+SOC band by solid red lines, and GW eigenvalues are shown by purple circles. The Fermi level is set to zero.*

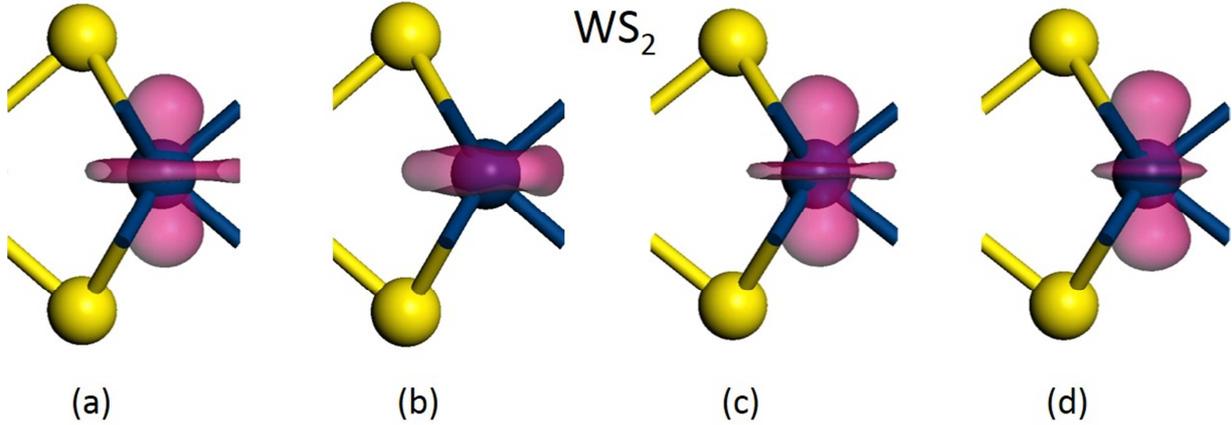

*Figure 2. Partial charge densities of (a) conduction band minimum (CBM) and (b) valence band maximum (VBM) states of 2D WS$_2$ without strain, and (c) CBM and (d) VBM states of 2D WS$_2$ under 5% strain. Yellow spheres represent sulfur atoms, and dark blue spheres represent tungsten atoms. All charge density iso-surfaces are shown at the same level of charge density.*

It is clear that all of these contributions need to be considered when carrying out a theoretical estimate of the band gaps values. In order to maintain closest structural correspondence with the actual systems, we used the 2H phase with experimental equilibrium room-temperature in-plane lattice constants of the bulk materials in all our calculations. The experimental lattice constants that were used were *a*=3.161 Å for MoS$_2$, [14] and *a*=3.153 Å for WS$_2$. [15] Corresponding biaxial tensile strains were defined with respect to these values. Vertical positions of sulfur atoms were optimized at each value of the strain. It was verified that the discrepancy between these optimized and actual positions of sulfur atoms did not affect the band structure significantly. Figure 2 shows optimized atomic positions and partial charge densities of conduction band minimum (CBM) and valence band maximum (VBM) states of WS$_2$ under 0% and 5% strain. The CBM orbital exhibits $d_{z^2}$ character, while the VBM orbital has $d_{z^2} + d_{x^2-y^2}$ character, in both strained and unstrained cases. For MoS$_2$, all corresponding orbitals are similar to those of WS$_2$.

Table I summarizes lattice constants, polarizabilities, exciton and trion binding energies, as well as corrections to the PBE band gaps given by the SOC and GW terms for monolayer MoS$_2$ and WS$_2$ under different biaxial strains. The in-plane polarizability $\chi_{2D}$ was calculated according to Eq. (2) from the DFT dielectric function obtained using the random phase approximation and neglecting local field effects. By separating the total value of the band gap into contributions from various terms, as presented in Table I, one can appreciate the influence of various effects on the magnitude of the gap.

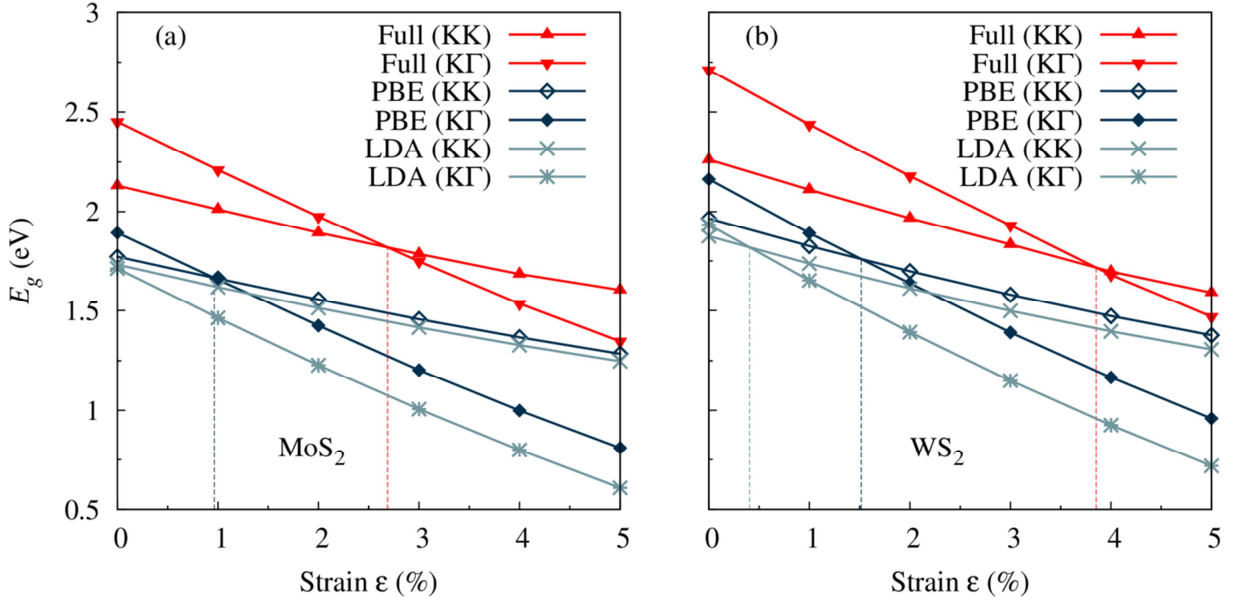

*Figure 3. Band gaps of monolayer MoS$_2$ and WS$_2$ under different biaxial strains in various approximations. 'Full' designates values of the band gaps calculated according to Eq. (5). Linear lest-square fits $E_g = a\varepsilon + b$ have been carried out for the full results, yielding the following parameters: $E_g = -0.106\varepsilon + 2.117$ for the MoS$_2$ direct gap, $E_g = -0.222\varepsilon + 2.433$ for the MoS$_2$ indirect gap, $E_g = -0.135\varepsilon + 2.248$ For the WS$_2$ direct band gap, and $E_g = -0.249\varepsilon + 2.692$ for the WS$_2$ indirect gap.*

The final direct and indirect gaps are shown in Fig. 3. For comparison, the PBE and LDA gaps are also shown. The final gaps $E_g$ were calculated according to

$$E_g = E_g(PBE) + \Delta E_g^{GW} + \Delta E_g^{SOC} + E_{ex} + E_{tri} \qquad (5)$$

In Eq. (5), $E_g(PBE)$ is the PBE band gap, $\Delta E_g^{GW}$ is the GW correction to it, $\Delta E_g^{SOC}$ is the change of the PBE gap after the addition of the spin-orbit coupling, $E_{ex}$ is the binding energy of exciton, and $E_{tri}$ is the binding energy of trion (relative to exciton). Note that while both trionic and excitonic peaks can usually be observed in experiment, the optical gap in the Eq. (5) is defined as the lowest-energy transition, and thus includes the energy of the trion. The zero-strain gap in MoS$_2$ calculated according to Eq. (5) is 2.13 eV. This theoretical value should be compared with the experimental gap of 1.85 eV. [13] The calculated gap in WS$_2$ at zero strain is 2.26 eV, as compared to 1.99 eV experimental gap. [13] In both cases, the calculated gaps are 0.27 or 0.28 eV larger than the experimental ones. This difference may be due to the yet unaccounted electron-phonon band gap renormalization. The electron-phonon renormalization decreases the gap, the magnitude of the decrease reaching 0.6 eV in some materials such as diamond. [16] The direct to indirect gap transitions are predicted to be at strains of -0.14% and 0.96% in MoS$_2$, by LDA and PBE functionals, correspondingly. After taking into account corrections from Eq. (5), the predicted transition strain value becomes 2.69%. Similarly, in WS$_2$, LDA and PBE functionals yield values of 0.41% and 1.52%, whereas full calculations predicts the transition at 3.85%.

Since the gaps change almost linearly with strain, linear least-square fitting can be performed, giving the following expressions for the gap variations: $E_g = -0.106\varepsilon + 2.117$ for the MoS$_2$ direct gap, $E_g = -0.222\varepsilon + 2.433$ for the MoS$_2$ direct gap, $E_g = -0.135\varepsilon + 2.248$ for the WS$_2$ direct gap, and $E_g = -0.249\varepsilon + 2.692$ for the WS$_2$ indirect gap. It is clear from Fig. 3 that while the corrections significantly change the gap value, its strain dependence $dE_g/d\varepsilon$ remains almost unchanged and is reproduced at the very basic DFT level very well. This general observation suggests significant reduction in the volume of computations needed to determine the direct-indirect transition strain: only zero strain value $E_g(\varepsilon = 0)$ must be computed with all effects accounted for, while the slopes can be assessed with inexpensive PBE.

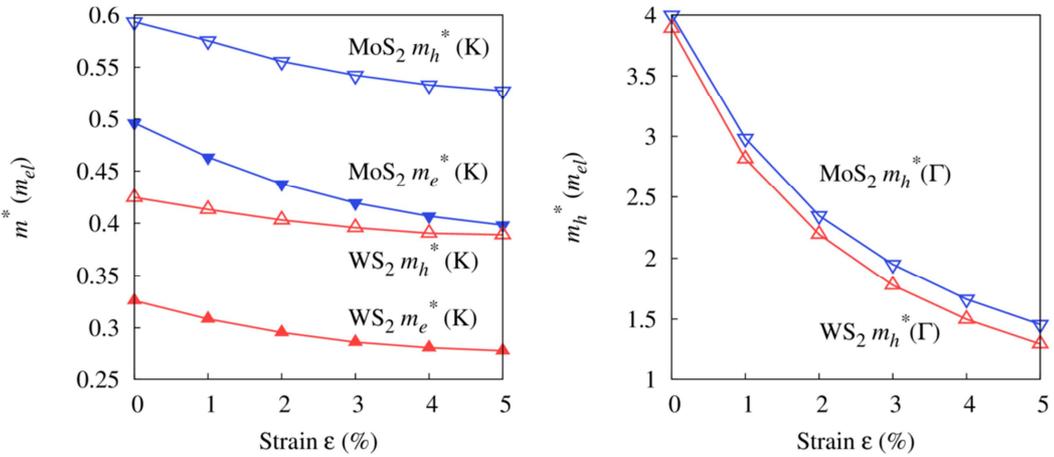

Figure 4. Effective mass of (a) electron and (b) hole for monolayer WS$_2$ under different biaxial strains calculated with PBE.

The effective masses of carriers that were used in finding the exciton and trion binding energies were calculated by fitting the band extrema to a parabola according to $\mathrm{E} = \hbar^2 k^2/2m_e m^*$, where $m_e$ is the electron mass in vacuum. A k-point spacing of 0.05 Å$^{-1}$ was used. Electron and hole effective masses ($m^*$) at different strains are shown in Fig. 4. The effective mass of MoS$_2$ is higher than that of WS$_2$, and unstrained hole effective mass at Γ point is about 8 times of that at K point for both materials. For both materials, electron and hole effective masses at K point and Γ point all decrease as the strain increases, the rate of reduction of the hole effective mass at Γ point being much higher than those for both carriers at the K point.

In conclusion, we have considered the corrections given by the GW approximation, spin-orbit coupling, exciton and trion binding energies to the Kohn-Sham eigenvalues of strained monolayer MoS$_2$ and WS$_2$ to accurately predict the band gaps and direct to indirect gap transition strains in these materials. Our theoretically predicted gaps are 0.27 or 0.28 eV larger than those measured experimentally, probably due to our neglecting the electron-phonon band gap renormalization. We predict the exact direct-indirect gap transition including all effects at 2.69% in MoS$_2$ and 3.85% in WS$_2$.

**Acknowledgements** The authors thank Evgeni Penev for stimulating discussions. This work was supported by the U.S. Army Research Office MURI grant W911NF-11-1-0362, and in part (for L.W.) by the Robert Welch Foundation(C-1590). Computer resources were provided by XSEDE, which is supported by NSF grant OCI-1053575, under allocation TG-DMR100029; and the DAVinCI cluster acquired with funds from NSF grant OCI-0959097.

TABLE I. Lattice constants ($a$), polarizibality ($\chi_{2D}$), the exciton binding energy ($E_{ex}$) and the trion binding energy ($E_{tri}$), corrections to the PBE band gaps given by the GW ($\Delta E_g^{GW}$) and SOC terms ($\Delta E_g^{SOC}$) for monolayer MoS$_2$ and WS$_2$ at different biaxial strains.

| | strain | $a$ (Å) | $\chi_{2D}$ | $E_{ex}$ (meV) | | $E_{tri}$ (meV) | | $\Delta E_g^{GW}$ (eV) | | $\Delta E_g^{SOC}$ (meV) | |
|---|---|---|---|---|---|---|---|---|---|---|---|
| | | | | KK | KΓ | KK | KΓ | KK | KΓ | KK | KΓ |
| MoS$_2$ | 0% | 3.160 | 6.989 | -532 | -600 | -24 | -26 | 1.002 | 1.209 | -78 | -9 |
| | 1% | 3.192 | 7.170 | -515 | -575 | -24 | -25 | 0.978 | 1.180 | -79 | -9 |
| | 2% | 3.223 | 7.286 | -502 | -557 | -23 | -25 | 0.956 | 1.152 | -79 | -8 |
| | 3% | 3.255 | 7.485 | -488 | -537 | -23 | -24 | 0.933 | 1.126 | -80 | -7 |
| | 4% | 3.286 | 7.678 | -476 | -520 | -22 | -23 | 0.902 | 1.093 | -80 | -6 |
| | 5% | 3.318 | 8.023 | -459 | -498 | -21 | -22 | 0.889 | 1.073 | -81 | -6 |
| WS$_2$ | 0% | 3.153 | 6.462 | -508 | -579 | -25 | -27 | 1.104 | 1.252 | -270 | -87 |
| | 1% | 3.185 | 6.519 | -498 | -564 | -24 | -26 | 1.086 | 1.228 | -266 | -72 |
| | 2% | 3.216 | 6.587 | -490 | -550 | -24 | -26 | 1.059 | 1.197 | -264 | -61 |
| | 3% | 3.248 | 6.797 | -475 | -530 | -23 | -25 | 1.030 | 1.162 | -263 | -51 |
| | 4% | 3.279 | 6.800 | -473 | -524 | -23 | -25 | 0.993 | 1.114 | -263 | -44 |
| | 5% | 3.311 | 6.953 | -464 | -511 | -23 | -24 | 0.966 | 1.085 | -263 | -37 |